# Ultra-Compact Coupling Structures for Heterogeneously Integrated Silicon Lasers

An He, Lu Sun, Hongwei Wang, Xuhan Guo, Yikai Su

*Abstract*—Due to the inherent in-direct bandgap nature of Silicon, heterogeneous integration of semiconductor lasers on Silicon on Insulator (SOI) is crucial for next-generation on-chip optical interconnects. Compact, high-efficient and high-tolerant couplers between III-V light source and silicon chips have been the challenge for photonic integrated circuit (PIC). Here, we redesign the taper adiabatic coupler with the total coupling length of only 4 µm, and propose another two novel slot coupler and bridge-SWG coupler with both coupling length of 7 µm, to heterogeneously integrate III-V lasers and silicon chips. We study theoretically the optical mode coupling process through the redesigned taper coupler, the final coupling results match well with the simulation in 3D-FDTD. The three compact couplers represent fundamental TE mode coupling efficiencies all over 90%, even 95.7% for bridge-SWG coupler, to the best of our knowledge, are also the shortest coupling structures (7 um). Moreover, these coupling structures also possess excellent fabrication tolerance.

*Index Terms*—Heterogeneous integration, photonic integrated circuits, semiconductor lasers, optical interconnections.

## I. INTRODUCTION

By exploiting mature and standard microelectronics CMOS foundries, Si photonics have been booming due to the promise of low-cost fabrication, low power consumption and compact circuits that integrate photonic and microelectronic elements, with rapidly commoditized Si photonic integrated circuit (PICs) in information and communication technology [1]–[5]. Nowadays as the convergence of optics and electronics at the chip level is becoming a necessity for the next-generation processors and data communications, the high dense intra- and inter-chip interconnections in Si photonics are drawing enormous attention [6]–[10]. However due to their indirect bandgap, full-scale deployment of Si photonics is still hampered by the lack of directly integrated light sources. Intensive research efforts have been devoted to the realization of integrating electrically-pumped III-V lasers in Si photonics and there are three main approaches namely flip-chip integration [11]–[14], epitaxial growth [15]–[19], and bonding technology [20]–[24].

Directly mounting pre-fabricated III-V lasers using flip-chip technology is currently preferred by the industry. It allows the pre-selection of known-good lasers, but the limited alignment tolerance and high packaging cost make it difficult for further scaling. With some recently exciting results [25]–[27], monolithically integrated on-chip light sources by epitaxial growth may realize the high-density integration of lasers in Si photonics economically, which might be one of the ultimate goal. However, due to the large mismatch in lattice constant (i.e. 8% for InP/Si and 4% for GaAs/Si), it is still in the early stages of research and the fabrication process is not mature, the material quality and laser reliability still need improvement. What's more, the necessity of thick buffer filter layers and methods for laser-chip light coupling is still open to discuss. The bonding technology is to transfer sheets of epitaxial III-V material to the Si platform (on the III-V die or III-V wafer level) and to process the III-V opto-electronic components afterwards. This makes the integration process much easier since no stringent alignment is required and it allows exploiting lithography to align the III-V components to the underlying SOI waveguide circuit on a wafer level. Therefore, bonding-based heterogeneous integration has been widely explored in the past decade and has proven to be the most successful approach up to now for dense III-V laser integration on Si [2], [28]–[30]. For bonding technology, the divinylsiloxane-bis-

The authors are with the State Key Laboratory of Advanced Optical Communication Systems and Networks, Department of Electronic Engineering, Shanghai Jiao Tong University, Shanghai 200240, China (e-mail: he--an@sjtu.edu.cn; sunlu@sjtu.edu.cn; 018034210002@sjtu.edu.cn; guoxuhan@sjtu.edu.cn; yikaisu@sjtu.edu.cn), corresponding author: Xuhan Guo benzocyclobutene (BCB), as an adhesive bonder, has been widely implemented to combine various laser sources onto Si chip [31]–[34] due to its easy procedures, high stability, high bonding strength and void-free bonds [35].

Among various Si/III–V heterogeneously integrated devices, a common critical issue is how to design a compact mode coupler structure to route light efficiently between the III–V active waveguide and the Si waveguide. The tapered silicon coupler has been widely used to couple light from III-V laser source to strip Si waveguide [20], [31], [32], [34], [36], but the relative long coupling length (CL) from tens to hundreds microns hampered the integration level and miniaturization of PIC. In order to densely integrate the III–V semiconductors with the Si PICs, we propose two novel coupler structures based on slot and subwavelength grating (SWG) waveguides.

Due to the unique ability to significantly confine and guide light in the low-index slot rather than the high index Si core [37], slot waveguides break the material limitation and the structure closure of traditional solid-core waveguides and provide a convenient way for interaction between the guided light and the filling materials [38]. Various functional devices based on slot waveguides have been fabricated, such as Si photonics bio-sensing [39], polarization beam splitter [40], and modulators [41]. As a periodic waveguide, SWG is highly dependent on the operating frequency, which needs to be set in certain region, or the light will be reflected or radiated [42], [43]. And because of the ability to easily modulate effective index and accurately control the distribution of the electromagnetic field, SWG has significant application in fiber-chip coupler [44], optical filter [45], multiplexer [46] and demultiplexer [47]. To reduce the effective index mismatch between SWG and conventional strip waveguide, achieving the adiabatically conversion from the SWG waveguide with Bloch–Floquet mode to the strip waveguide with fundamental mode, a bridge-SWG with a narrower strip or taper Si bridge imbedded in SWG was proposed [48]–[50]. However, none of the slot Si waveguide and bridge-SWG Si waveguide structures have been studied for heterogeneously integration couplers before.

In this paper, in order to introduce the novel coupler structures and for a fair comparison, we first investigate and calculate the scattering loss caused by the sidewall roughness of strip waveguide, slot waveguide and SWG waveguide, which reduces the light transmission efficiency [51]–[53]. Secondly, we analyze and calculate the dynamic transmission of optical power during coupling process from silicon strip waveguide to III-V laser through a taper coupler by theory and 3D finite-difference time domain (3D-FDTD) simulation, and redesign an ultra-compact taper coupler for integrating III-V laser to silicon strip waveguide. Thirdly, we propose two novel coupling structures, i.e. slot coupler and bridge-SWG coupler, for guiding light from Si waveguide to III-V laser. We also study the influence of fabrication tolerance of these three couplers on the coupling efficiency and give the state-of-the-art heterogeneous integration coupler structures and performance comparison before the conclusion.

## II. SCATTERING LOSS OF STRIP, SLOT AND BRIDGE-SWG WAVEGUIDES

Achieving the low optical transmission loss is one of the main challenges of the silicon-based waveguide products [54]. However, in the fabrication process of waveguide, the photo-lithographic will produce surface roughness inevitably, which causes the optical losses and performance degradation [3]. In the last decades, enormous studies have been devoted to the waveguide sidewall surfaces caused optical scattering losses [51], [52], [55]. So far only a few efforts have been aimed at the direct comparison among strip, slot and bridge-SWG waveguides for light coupling [56].

As shown in Figure 1(a), in the strip waveguide, the optical field is confined in the core region due to the total internal reflection. But the light can be distributed in low index area, such as silica, in slot and bridge-SWG waveguides [Figure 1(b) and (c)], this unique ability has attracted significant interest due to various application as mentioned above. However, a high sensitivity to sidewall roughness σ induced scattering loss in these structures compared with strip waveguides casts doubt on their performance. Thus, investigating the scattering loss of these three waveguides before the coupler structure design is of high significance.

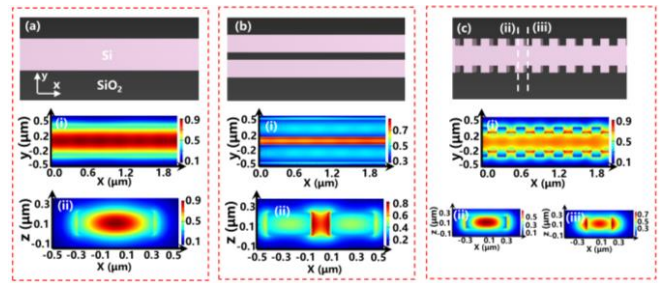

Fig. 1. Top view of silicon-based (a) strip, (b) slot and (c) Bridge-SWG waveguide. The total widths of these three waveguides are 600 nm, the slot width of slot waveguide is 150nm, the period and DC of the bridge-SWG

waveguide are 300 nm and 0.5, respectively. Mode profiles of the propagated fundamental TE (quasi-TE) mode in (a-i) strip, (b-i) slot and (c-i) bridge-SWG waveguide. Cross-sectional view of electric field distribution in (a-ii) strip waveguide, (b-ii) slot waveguide, (c-ii) silicon block and (c-iii) strip-bridge of bridge-SWG waveguide.

The physics underlying scattering loss roots in sidewall-roughness lies in the local variations of the waveguide width, induces the variation of the waveguide refractive index accordingly, and consequently causes the appearance of scattering loss [57]. In this paper, we calculate silicon waveguides depicted in Figure 1 with $SiO_2$ cladding, which fully fills the interspaces in slot waveguide and bridge-SWG waveguide. The top and bottom interfaces in commercial well-fabricated SOI wafers are nearly atomically flat and do not contribute significantly to scattering loss [58]. Thus, only the sidewall roughness is considered. Based on this model, the scattering loss coefficient α can be elicited from Payne and Lacey [59]:

$$\alpha = \frac{\sigma^2}{\sqrt{2}k_0 d^4 n_1} f \cdot g \quad (1)$$

where, $f$ and $g$ can be expressed as:

$$f = \frac{x\{[(1+x^2)^2 + 2x^2 y^2]^{1/2} + 1 - x^2\}^{1/2}}{[(1+x^2)^2 + 2x^2 y^2]^{1/2}} \quad (2)$$

$$g = \frac{U^2 V^2}{1+W} \quad (3)$$

$$U = d\sqrt{n_1^2 k_0^2 - \beta^2} \quad (4)$$

$$V = k_0 d\sqrt{n_1^2 - n_2^2} \quad (5)$$

$$W = d\sqrt{\beta^2 - n_2^2 k_0^2} \quad (6)$$

$$x = W\frac{L_c}{d}, \quad y = \frac{n_2 V}{n_1 W \sqrt{\Delta}},$$
$$\Delta = \frac{n_1^2 - n_2^2}{2n_1^2} \quad (7)$$

Here, $d$ is the half width of waveguide. Core and cladding refractive indexes are denoted as $n_1$ and $n_2$, respectively, $k_0 = \omega/c$ is the free-space wavenumber for an angular frequency $\omega$ and the vacuum speed of light $c$, the correlation length of the surface roughness $L_c$ and σ are set as 100 nm and 1nm [56], respectively, $\beta$ is the light propagation constant. From this theory, the refractive index is main factor affecting the scattering loss.

In the following studies, we compare the scattering loss of the three kinds of Si waveguides, i.e. strip waveguide, slot waveguide and bridge-SWG waveguide. Figure 1(a)-(c) depict a SOI wafer to form various silicon waveguides. The SOI wafer considered here has a standard 220 nm top silicon layer and 2 μm $SiO_2$ substrate and cladding. To give a relatively fair comparison, the total silicon width of slot waveguide and bridge-SWG waveguide is set to 600 nm. The period and bridge width of bridge-SWG waveguide are set to 300 nm and 350 nm, respectively. TE polarization light is considered only. The influence of strip width for strip waveguide, slot width for slot waveguide and duty cycle (DC) for bridge-SWG waveguide on scattering loss is analyzed respectively.

The theoretically calculated results are demonstrated in Figure 2. For the strip waveguide, the scattering loss decreases as the waveguide width increases. As the waveguide broadens, the optical field tends to gather at the center of the waveguide, away from the rough side wall. This field distribution causes the inappreciable influence of sidewall roughness on the scattering loss. For the slot waveguide, the increase of the slot width causes the reduction of scattering loss. The extension of slot width generates a larger optical field volume, resulting in the attenuation of the interaction between side wall and the field. Inversely, for the bridge-SWG waveguide, the optical field volume shrinks as the DC increases, causing the enhanced interaction between light and sidewall roughness with increasing scattering loss. And the results indicate that compared with 600 nm wide strip waveguide, slot (slot width below 130 nm) and bridge-SWG waveguides possess increased scattering losses.

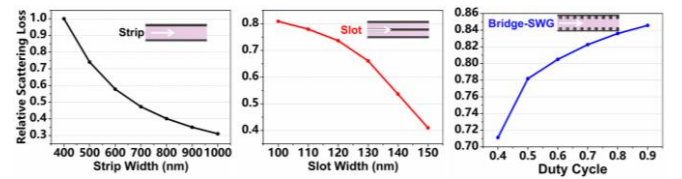

Fig. 2. Normalized scattering loss coefficient α/max ($\alpha_{strip}$) of strip, slot and bridge-SWG waveguide. Max ($\alpha_{strip}$) represents the maximum of the scattering loss coefficient in strip waveguide.

III. HETEROGENEOUS INTEGRATION STRUCTURE

High integration density is always pursued in Si photonics. A compact coupler can observably shrink the footprint of Si/III–V heterogeneously integrated devices.

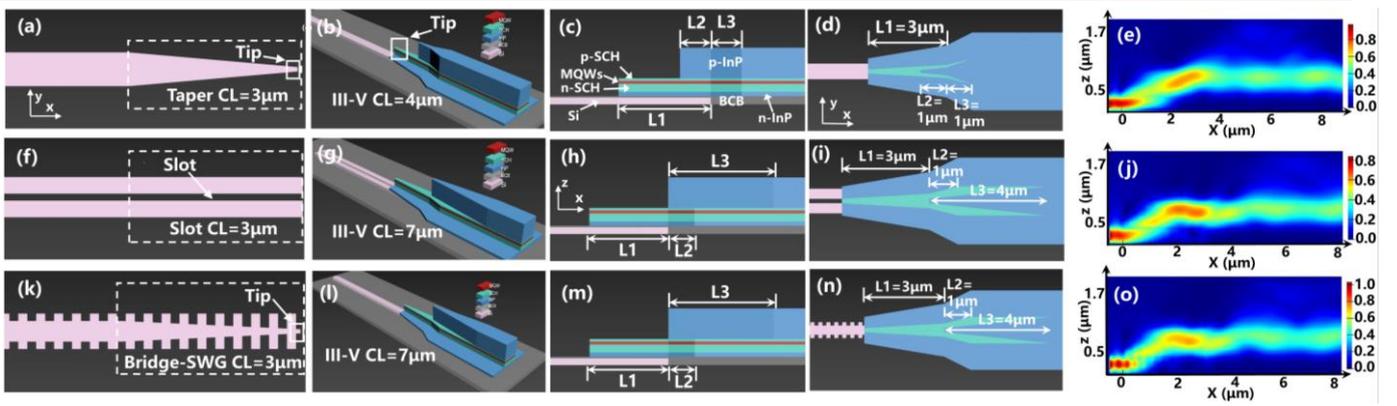

Figure.3. Schematics of heterogeneous integration of Si waveguides and III-V laser sources through (a) - (d) taper coupler, (f) - (i) slot coupler and (k) - (n) bridge-SWG coupler. (e) Mode transformation from Si taper waveguide to III-V laser, the coupling range is from 0 μm to 4 μm. (j), (o) Mode transformation from slot and bridge-SWG Si waveguides to III-V lasers, respectively, the coupling ranges are both from 0 μm to 7 μm.

Based on the scattering loss study, to integrate III-V lasers on strip, slot and bridge-SWG waveguide, we redesign and optimize the taper coupler with entire CL of only 4 μm and propose two novel slot and bridge-SWG couplers with entire CL of 7 μm for heterogeneous integration through 50 nm thick BCB (Figure 3), which can be realized by the existing BCB bonding technology [60]–[62]. We first calculate the theoretical coupling process for taper coupler, then study the coupling efficiency of total mode and fundamental TE (quasi-TE) mode of these three couplers. Also, the fabrication tolerance is investigated. To reduce the CL, the III-V multi-step taper couplers consisting of three independent sections is implemented [21], [22]. The thickness and refractive index in each layer are listed in Table I [22].

TABLE I.
SPECIFICATIONS FOR THE THICKNESS AND REFRACTIVE INDEX IN EACH LAYER

| Layer | Thickness (μm) | Refractive index |
| --- | --- | --- |
| Si | 0.22 | 3.477 |
| n-InP | 0.14 | 3.17 |
| n-SCH | 0.26 | 3.46 |
| MQW | 0.1 | 3.52 |
| p-SCH | 0.08 | 3.46 |
| p-InP | 1 | 3.17 |
| BCB | 0.05 | 1.543 |

Figures 3 shows the schematic structures of the proposed tri-sectional taper, slot and bridge-SWG coupler. The first section is a vertical coupler for light coupling from the Si waveguide to a 0.44 μm thick III–V waveguide, including n-InP layer, n-SCH (separate confinement heterostructure) layer, multiple quantum well (MQW) layer, and p-SCH layer. The second section is a taper for light coupling to a wider III–V waveguide. The third section is a taper for light transferring to the III–V waveguide with the p-InP layer. This 3D tri-sectional taper not only induces low reflection, similar to multilevel tapered couplers [63], but also avoids exciting high-order modes in the III–V waveguide [21].

A. *Ultra-compact Taper Coupler*

Due to the low loss, and high fabrication tolerance, most integration approaches employ taper coupler as mode converter [20], [31], [32], [34], [36]. One way to reduce the CL is to use a multistep or complex shape tapered structure in the III–V waveguide [21], [22].

In our design, the multistep taper is chosen as shown in Figure 3(a)-(d). The optimized III-V tapered coupler consists of three sections. The first section (L1) is a 3 μm vertical coupler, and partially covered with a 1 μm tapered p-InP layer (L2) which is the second section. The third section is a 1 μm long tapered III-V materials (L3).

We theoretically analyze the optical mode coupling process between Si taper and III-V first section without p-InP layer, which determines the final coupling efficiency. The Si coupler width shrinks from 600 nm to 150 nm at the length of 3 μm. Upon the Si taper, the width of n-InP taper increases from 1 μm to 2 μm, and the widths of tapered SCH and MQW layers are all from 150 nm to 500 nm. Propagation of optical field in a perturbed dielectric structure can be described by the coupled mode theory. According to the perturbation theory, when the gap between two waveguides is small enough, one waveguide mode can be coupled into another

$$\frac{dA}{dz} = -i\kappa_{ab} B e^{i2\delta z}$$

$$\frac{dB}{dz} = -i\kappa_{ba} A e^{-i2\delta z}$$

$$2\delta = (\beta_a + k_{aa}) - (\beta_b + k_{bb})$$

Where $A$, $B$ and $\beta_a$, $\beta_b$ are the optical mode amplitudes and propagation constants of waveguide a and b, respectively. $\kappa_{ab}$ and $\kappa_{ba}$ represent the exchange coupling coefficients between waveguide a and b. which is defined as:

$$k_{ab} = \frac{\omega}{4}\varepsilon_0 \iint E_a^* \Delta n_a^2(x,y) E_b dxdy$$

$$k_{ba} = \frac{\omega}{4}\varepsilon_0 \iint E_b^* \Delta n_b^2(x,y) E_a dxdy$$

$$k_{aa} = \frac{\omega}{4}\varepsilon_0 \iint E_a^* \Delta n_b^2(x,y) E_a dxdy$$

$$k_{bb} = \frac{\omega}{4}\varepsilon_0 \iint E_b^* \Delta n_a^2(x,y) E_a dxdy$$

$\omega$ is the angular frequency, $\varepsilon_0$ is the vacuum permittivity, $E_a$ and $E_b$ are the electric field profiles of waveguide a and b, respectively, $\Delta n_a(x,y)$ and $\Delta n_b(x,y)$ are the periodic perturbation in the waveguide a and b.

The theoretically calculated results are exhibited in Figure 4 as the guideline for the taper design. As the light travels forward from Si strip waveguide, it gradually couples into the III-V materials through the taper coupler. This coupling process is verified by simulation in 3D-FDTD. During the coupling process, a small portion of TE1 mode is excited and most is converted to TE0 mode. This excited TE1 mode will be furtherly coupled to TE0 mode through the second section of III-V taper. The final coupling results match well with the theoretical results, but the process of mode transition is a little different. Compared with the ultra-compact taper coupler proposed by Huang [21] which is 8 μm, we reduce the length of first section acting as vertical coupling, and adjust the position of p-InP layer. The entire CL of this redesigned taper coupler is only 4 μm, which is half of [21], to the best of our knowledge, is the shortest. This theoretical analysis can also be applied to the other two coupler structures under study.

mode due to overlapped evanescent wave, and the amplitude of each waveguide mode along the propagation direction is determined by a set of differential equations [64]:

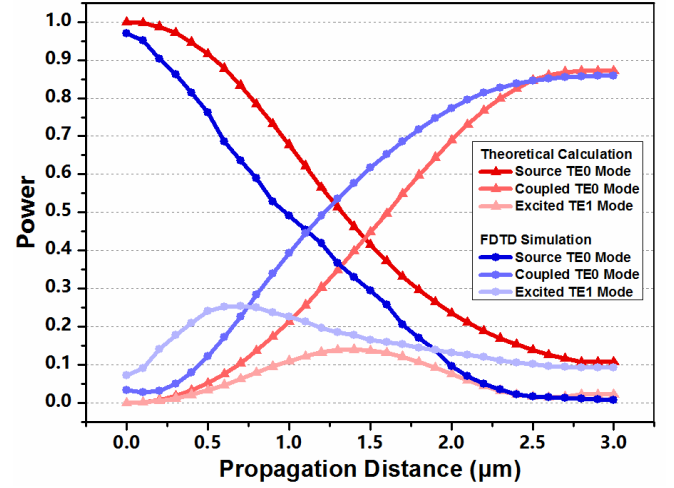

Fig. 4. The change of optical power during coupling process through taper coupler calculated by theory and simulated by FDTD.

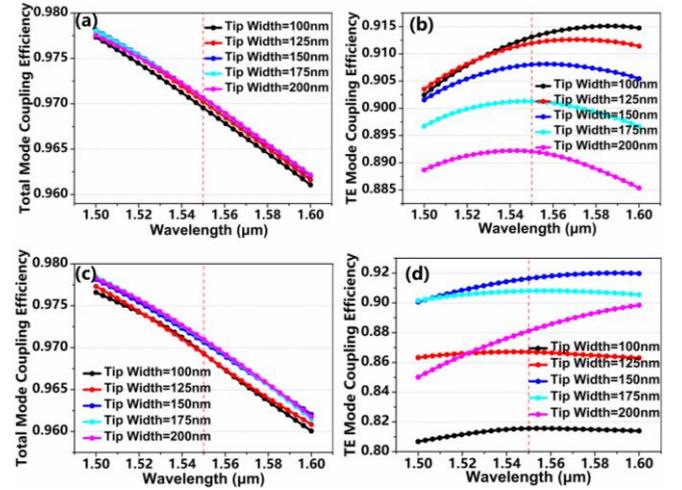

Fig. 5. Coupling performance and fabrication tolerance of the taper coupler. The influence of the tip width of Si taper on (a) total mode and (b) fundamental TE mode coupling efficiency, the tip width of the first section III-V taper is 150 nm. The influence of the tip width of the first section III-V taper on (c) total mode and (d) fundamental TE mode coupling efficiency, the Si tip width is 150 nm.

From Figure 5(a) and (b), one can see that when the tip width of Si taper is 100 nm, the total mode and fundamental TE mode coupling efficiency can be maintained over 96% and 90% with a wide bandwidth from the wavelength of 1500 nm to 1600 nm, respectively. At 1550 nm, the total mode and fundamental TE mode coupling efficiencies are 97.1% and 91.3%. The fabrication tolerance of the tip width of Si taper and the first section III-V taper is also analyzed, we found that the tip width of the Si taper barely affects the total mode and fundamental TE mode coupling efficiency [Figure 5(a) and (b)]. The tip width increases from 100 nm to 200 nm, the decreasing is only 0.1% for total mode coupling efficiency and 2.1% for TE mode coupling efficiency. This 100 nm tolerance is far greater than the deviation of fabrication technology. As presented in Figure 5(c), for the III-V taper in the first section, the total mode coupling efficiency fluctuates within 0.5% when the tip width changes from 100 nm to 200 nm, which indicates the impact of tip width on total mode coupling efficiency can also be neglected. While the optimized TE mode coupling efficiency is obtained at the tip width of 150 nm at the wavelength of 1550 nm [Figure 5(d)].

*B. Slot and Bridge-SWG Couplers*

In this section, two novel coupling structures, slot and bridge-SWG couplers, are studied. We keep the coupling region of slot structure the same during light propagation, as shown in Figure 3(f) and (g). While for bridge-SWG structure, in the Si coupler, the bridge strip shrinks to a tip along the light propagating direction [Figure 3 (k) and (l)]. The structure and dimension of III-V materials of slot and bridge-SWG coupler consist of the same tri-section structure, which are inspired from the taper coupler and optimized in 3D-FDTD. The first section is a 3 μm long vertical tapered coupler for light coupling from the silicon waveguide to III–V waveguides without the p-InP layer. The second section is a 1 μm long taper for light coupling to a wider III–V waveguide. The third section is a 4 μm long tapered p-InP layer covering the second section and final III-V mesa. Thus, the entire CLs of these two novel couplers are both 7 μm.

From Figure 6(a), the slot width of the slot waveguide has little influence on the total mode coupling efficiency, but the TE mode coupling efficiency is affected obviously [Figure 6(b)]. The highest TE mode coupling efficiency is at the slot width of 100 nm. But considering the fabrication limitation and stability, the slot width of 150 nm may be advisable. As presented in Figure 6(c), the optimal total mode coupling efficiency appears at the tip width of 150 nm in the first III-V section, and the TE mode coupling efficiency is similar at the tip width of 100, 125 and 150 nm [Figure 6(d)], thus the tip width of 150 nm can be chosen due to the fabrication technology.

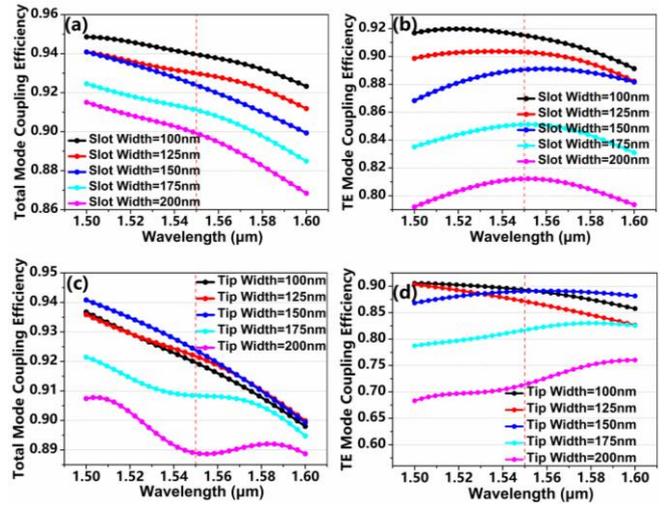

Fig. 6. Coupling performance and fabrication tolerance of the slot coupler. The influence of the slot width on (a) total mode and (b) fundamental TE mode coupling efficiency, the tip width of the first section III-V taper is 150 nm. The influence of tip width of the first section III-V taper on (c) total mode and (d) fundamental TE mode coupling efficiency, the Si slot width is 150 nm.

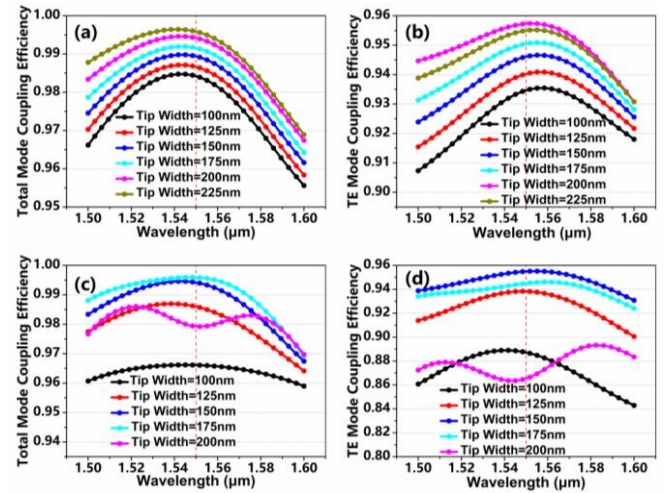

Fig. 7. Coupling performance and fabrication tolerance of the bridge-SWG coupler. The influence of the slot width on (a) total mode and (b) fundamental TE mode coupling efficiency, the tip width of the first section III-V taper is 150 nm. The influence of tip width of the first section III-V taper on (c) total mode and (d) fundamental TE mode coupling efficiency, the Si tip width is 150 nm.

From Figure 7(a), bridge-SWG coupler exhibits an ultra-high total mode coupling efficiency, close to 100% at the wavelength of 1550 nm, and the fundamental TE mode coupling efficiency is also higher than taper and slot coupler. The tip width of the tapered bridge has negligible influence on the total mode coupling efficiency, the highest TE mode coupling efficiency is at the tip width of 200 nm [Figure 7(b)].

Also, the total mode coupling efficiency is barely affected by the tip width of the III-V taper, as shown in Figure 7(c). But as presented in Figure 7(d), the optimal TE mode coupling efficiency appears at the tip width of 150 nm in the first section, thus the tip width of the III-V taper in the first section can be 150 nm. Therefore, the tip width of the III-V in the first section for the three coupling strategies can be all chosen as 150 nm. And the bridge-SWG can achieve the optimal TE mode coupling efficiency within the fabrication resolution.

In addition, the comparison between the three couplers is investigated in this paper and other state-of-the-art couplers are given in table II. To our best knowledge, we proposed the most compact coupling structures, and the experimental verification is on the way.

TABLE II.

COMPARISON OF THE COUPLER STRUCTURES. THE FIRST THREE LINES REPRESENT THE TAPER, SLOT AND BRIDGE-SWG COUPLER INVESTIGATED IN THIS PAPER. THE # AND * INDICATE SIMULATION AND EXPERIMENT RESULTS. THE EFFICIENCY REPRESENTS THE SIMULATION RESULTS.

| Coupler structure | CL (μm) | Si Thickness (nm) | Platform | Efficiency |
|---|---|---|---|---|
| **Taper** | **4#** | **220** | **SJTU** | **97%** |
| **Slot** | **7#** | **220** | **SJTU** | **94%** |
| **Bridge-SWG** | **7#** | **220** | **SJTU** | **99%** |
| Taper [54] | 25# | 220 | A*STAR | 97% |
| Taper [21] | 8# | 220 | ZJU | 95% |
| Taper [55] | 250# | NA | IMEC | 100% |
| Taper [56] | 120* | 500 | CEA-LETI | 97% |
| Taper [20] | 230* | 400 | IMEC | NA |
| Taper [34] | 215* | 400 | IMEC | NA |
| Trident [57] | 150* | 220 | IMEC | NA |
| Taper [12] | 50* | 750 | UCSB | NA |
| Taper [58] | 30* | 220 | A*STAR | 90% |
| Taper [59] | 200* | 500 | UCSB | NA |

IV. CONCLUSION

In this paper, an ultra-compact taper coupler is redesigned, two novel couplers, slot coupler and bridge-SWG coupler are proposed for dense heterogeneous integration silicon lasers. These three coupling structures possess compactness, high efficiency and high tolerance at the same time, which are highly demanded for silicon PIC. The entire CL of taper coupler is 4 μm, and 7 μm for both slot and bridge-SWG couplers. For such short coupler, the fundamental TE mode coupling efficiency can still reach 91.3%, 91.5% and 95.7% for taper, slot and bridge-SWG coupler at the wavelength of 1550 nm, respectively. Especially for bridge-SWG coupler, the fundamental TE mode coupling efficiency can maintain higher than 93% in 100 nm wavelength range. These couplers are the most compact compared with other state-of-the-art couplers. Moreover, these ultra-compact couplers, which can be practically fabricated by existing technology, also exhibit excellent fabrication tolerances.

FUNDING

ACKNOWLEDGEMENT

The authors would like to thank the support of State Key Laboratory of Advanced Optical Communication Systems and Networks, Shanghai Jiao Tong University, China